\documentclass[useAMS,usenatbib]{mn2e}
\usepackage{psfig, epsf, epsfig}
\title[The lost dark matter]{The total mass and dark halo properties 
of the Small Magellanic Cloud}
\author[K. Bekki
and S.  Stanimirovi\'c]{Kenji Bekki${}^1$\thanks{E-mail:
bekki@phys.unsw.edu.au} and
Sne\v{z}ana  Stanimirovi\'c${}^2$ \\
       ${}^1$School of Physics, University of New South Wales,
              Sydney 2052, NSW, Australia\\
      ${}^2$
Department of Astronomy, 
University of Wisconsin Madison,
6520 Sterling Hall
475 N, Charter Street
Madison, WI 53706}

\begin{document}

\date{Accepted, Received 2005 February 20; in original form }

\pagerange{\pageref{firstpage}--\pageref{lastpage}} \pubyear{2005}

\maketitle

\label{firstpage}

\begin{abstract}
We discuss the total mass and dark halo properties of the 
Small Magellanic Cloud (SMC) for reasonable $V$-band stellar-to-mass-to-light
ratios ($M_{\rm s}/L_{\rm V}$) based on 
the high-resolution neutral hydrogen (HI) observations of the SMC.
We adopt both the Burkert 
and the NFW profiles
for the dark matter halo of the SMC in order to model 
the radial profile of the observed rotation curve.
We show that the mass ($M_{\rm dm}$) and the mean density  
(${\rho}_{\rm dm}$) of the dark matter  halo within the optical radius ($\sim 3$ kpc) of
the SMC 
can be significantly  low for $M_{\rm s}/L_{\rm V}=2-4$ reasonable
for a galaxy dominated by older stellar populations.
For example,
the estimated $M_{\rm dm}$ and ${\rho}_{\rm dm}$ are
$7.5 \times 10^7 {\rm M}_{\odot}$  and $6.7 \times 10^{-4} {\rm M}_{\odot}$ pc$^{-3}$,
respectively, 
for $M_{\rm s}/L_{\rm V} = 3.8$. 
The models with lower $M_{\rm s}/L_{\rm V}$ ($<1.0$)
can show higher $M_{\rm dm}$ and ${\rho}_{\rm dm}$
yet  have difficulties
in reproducing the inner rotation curve profile both for the Burkert and the
NFW profiles.
The  Burkert profile with  a larger
core radius ($>1$ kpc) (thus a low density) and a large mass
(${\rm M}_{\rm dm} > 3 \times 10^9 {\rm M}_{\odot}$)  can be better
fit to the observed slowly rising rotation curve
than the NFW one for a reasonable  $M_{\rm s}/L_{\rm V} \sim2$.
This means that
most of the dark halo mass can be initially located outside the optical radius
in the SMC and thus that
the dark halo would have already lost 
a significant fraction of its original 
mass due to the strong tidal interactions with the Galaxy
and possibly with the Large Magellanic Cloud (LMC).
We suggest that the dark mater halo of the SMC  is likely to have
the initial total mass and core radius as large as, or larger than,  $6.5 \times 10^9 {\rm M}_{\odot}$
and $3.2$ kpc, respectively. 
We discuss limitations of the present model in estimating the
total mass of the SMC.
\end{abstract}


\section{Introduction}

Observational studies of the neutral hydrogen (HI) gas in  the Magellanic Clouds (MCs) have
played a key role in the understanding of many aspects  of their formation, such as the formation
of the Magellanic Stream (MS; Mathewson et al. 1974; Putman et al. 1998),
gaseous kinematics of the Large Magellanic Cloud (LMC; Staveley-Smith et al.
2003),  and the formation of the Magellanic Bridge (MB; Muller et al.
2004). These observations furthermore have attracted much attention
from theoretical studies of the roles of LMC-Small Magellanic Cloud (SMC)-Galaxy interactions
in the formation of the MS (e.g., Murai \& Fujimoto 1980),
the importance of the last strong LMC-SMC interaction in 
the formation of the peculiar HI morphology (Bekki \& Chiba 2007),
and the origin of the bifurcated kinematics in the MB 
(e.g., Muller \& Bekki 2007). 
Recent observations of the structural and kinematical
properties of the HI gas in the SMC have provided valuable
information on the mass distribution  of the SMC
(Stanimirovi\'c et al.  2004, S04).

\begin{table*}
\centering
\begin{minipage}{185mm}
\caption{Parameter vales for the representative models.}
\begin{tabular}{cccccccc}
{Model
\footnote{The mass model for the dark matter halo of the SMC: 
``NDM'', ``B'', and ``NFW'' 
represent a model with no dark matter
(or with no specific dark matter profiles considered),  the Burkert profile for
the radial density profile of the dark matter halo,
and NFW one, respectively.}}
& {$i$ ($^{\circ}$)
\footnote{The inclination angles of the gas disk in the SMC.}}
& {$M_{\rm S}/L_{\rm V}$
\footnote{The $V$-band stellar-to-mass-to-light-ratio.}} 
& {$M_{\rm dm}$ (${\rm M}_{\odot}$)
\footnote{The total  mass of the dark matter halo
in units of $10^9 {\rm M}_{\odot}$.}}
& {$r_{\rm 0}$ (kpc)
\footnote{The scale radius in the Burkert profile.}}
& {$r_{\rm s}$ (kpc)
\footnote{The scale radius in the NFW profile.}}
& {$c$
\footnote{$c$ ($=r_{\rm vir}/r_{\rm s}$) parameter in the NFW profile.}}
& Comments \\
NDM1 &  40 & 3.8  & - &  - & - &  -  & without dark matter halo \\
NDM2 &  30 & 3.8  & - &  - & - &  -  & lower inclination angle \\
NDM3 &  40 & 2.3  & - &  - & - &  -  & lower $M_{\rm s}/L_{\rm v}$ \\
B1 &  40 & 2.3  & 6.5 & 3.2 & - &  -  & Burkert profile \\
B2 &  40 & 2.3  & 79.3 & 9.5 & - &  -  & \\
B3 &  40 & 0.6  & 79.3 & 9.5 & - &  -  & very low $M_{\rm s}/L_{\rm v}$\\
B4 &  40 & 1.0  & 79.3 & 9.5 & - &  -  & \\
NFW1 &  40 & 2.3  & 6.5 & - & 2.5 &  10  & NFW profile\\
NFW2 &  40 & 2.3  & 6.5 & - & 5.1 & 5  &  \\
NFW3 &  40 & 2.3  & 79.3 & - & 17.7 & 5  &  \\
\end{tabular}
\end{minipage}
\end{table*}

One of intriguing results by S04 is that the HI gas in the SMC has  rotational kinematics with
the maximum circular velocity ($V_{\rm c}$) of $\sim 60$ km s$^{-1}$.
This kinematics is in a striking contrast to the stellar kinematics
with no or little  rotation (e.g., Harris \& Zaritsky 2006),
which recently have led Bekki \& Chiba (2008) to suggest
that the SMC have experienced a major merger event long time ago.
Another intriguing result in S04 is that $V_{\rm c} \sim 60$  km s$^{-1}$,
corresponding to the total mass of $2.4 \times 10^9 {\rm M}_{\odot}$,
can be explained by the mass distribution of baryonic components
(i.g., gas and stars) without a dark matter halo for the central 3 kpc:
this result however depends strongly on the adopted stellar
distribution.
These properties certainly make the SMC a highly unusual  dwarf galaxy.
However, no detailed dynamical studies have been conducted 
to  explain the origin of such unusual dark-matter halo.
%whether the observation by S04 is really consistent with
%the scenario that the SMC has no or littler  dark matter halo.
 
The purpose of this paper is thus to discuss  the existence of 
dark matter within the central 3 kpc of the SMC, 
based on the detailed models for the mass distributions
of the dark matter halo and the baryonic components.
We derive the possible total mass ($M_{\rm dm}$)
and mean density (${\rho}_{\rm dm}$) of the dark matter halo 
within the optical radius of the SMC 
required to reproduce the observed HI rotation curve. 
We discuss whether the modeled distributions  of   the dark mater halo
in the SMC can be consistent with the observed slowly rising 
rotation curve.
We also discuss a possible initial  mass of the SMC 
before the interaction with the Galaxy and the LMC.

This paper focuses on implications and interpretation of observational
results presented in S04. We stress that the observed rotation curve of
the SMC depends on several adopted parameters:
inclination, position angle, the disk thickness and shape of the SMC. In
addition, possible radial gas-flows in the SMC, caused by the
LMC-SMC-Galaxy interactions, can result in non-circular motions of HI
gas
and make the estimate of the rotation curve even harder.
Some of the adopted parameters can not be directly determined by
observations,  because the 3D
HI structure remains observationally unclear for the SMC.
Since observational uncertainties involved in the derivation of the
rotation curve have already been discussed in S04, we just briefly
summarize them in this paper and explain limitations they impose on the
derivation of the dark matter halo. We also point out that the total
mass
of the SMC derived in S04 agrees well with the estimate based purely on
the stellar kinematics (Harris \& Zaritsky 2006).

\section{Possible total mass and mean density of the dark matter halo
from observations}

We first  derive
the possible mass and density of the dark matter halo
of the SMC from the observed rotation curve (S04) {\it without
using models for the dark matter} in \S 2.1 and \S 2.2.
We then try to find the density profile
of the halo that can fit well to the observed rotation curve
using  models predicted from previous theoretical
studies in \S 3.  
We discuss how uncertainties in the observed HI rotation curve (e.g. a
possible existence of radial gas inflow) affect our
results in Section \S 2.3.

\subsection{The observed rotation curve}

Before discussing the possible total mass and density of the dark matter halo of the SMC,
we firstly describe the observed radial profiles for the total mass  ($M$) 
and the rotational velocity ($V_{\rm c}$) of each component (e.g., gas and stars) within a 
distance ($R$) from the center of the SMC  based on the results by S04.
We here assume that the SMC has a stellar luminosity of $4.3 \times
10^8 {\rm L}_{\odot}$ (i.e., $M_{\rm V}=-16.76$ mag),
 which is significantly smaller than
that in S04: the rotation curve from stars in the present paper
is different from that in S04.

As described in S04, the tilted-ring algorithm was used to derive
the HI rotation curve. Keeping the kinematic center fixed, and 
assuming an inclination angle of 40 degrees, the final rotation curves for
the receding and approaching sides of the HI velocity field
were derived. As the line-of-sight velocity dispersion in the SMC is high
(mean value ~22 km s$^{-1}$), the asymmetric drift correction was applied,
resulting in the final rotation curve being higher by about 10 km/sec
for $R>0.5$ kpc than the observed one.
In addition, to derive the deprojected rotation curve due to the HI gas alone, an
exponential density law was assumed for the vertical HI disk distribution with
a scale height of 1 kpc. The deprojected rotation curve due to the stellar
component was derived from the  surface brightness  distribution at V-band.
For full details please see  S04. 
%The details of the procedures to derive $V_{\rm c}$  for stars and gas
%in the SMC
%are shown in S04, we here just briefly describe them.
Thus the circular velocity ($V_{\rm c}$) at a given radius is a combination
from dark matter ($V_{\rm c, dm}$),
stars ($V_{\rm c, s}$), and gas ($V_{\rm c,g}$).
%\begin{equation}
%{V_{\rm c}}^2={V_{\rm c, dm}}^2+{V_{\rm c, s}}^2
%+{V_{\rm c, g}}^2
%\end{equation}
Fig. 1 shows derived radial profiles of $M$ and $V_{\rm c}$ for the model with
the most plausible value for inclination ($i=40$), $M_{\rm s}/L_{\rm V}=3.8$,
and with no  dark matter (referred to as ``NDM'' model from now on).
%There can be  some observational uncertainties in determining
%the values of two key properties of $i$ and  $M_{\rm s}/L_{\rm V} $.
One source of uncertainty that has not been taken into account previously is
the inclination angle.
S04 estimated $i$ at each radius and found that the mean value 
of $i$ is 40$^{\circ} \pm 6^{\circ} $ within the central 2.3 kpc.
After this radius $i$ was found to rise slightly outward. 
%Since the observational error  ($\Delta i$) in $i$  at a given $R$
%is pretty small ($\Delta \la 5^{\circ}$), we have propagated
We have hence assumed the observational error ($\Delta i$) in $i$  (at a given $R$)
$\Delta = 5^{\circ}$ and propagated this value to estimate the uncertainty
in $M$ and $V_{\rm c}$  (as shown in Fig. 1).
%We however plot the observational error bars in  
%corresponding to $\delta =i \pm 5^{\circ}$ at given radii
%in order to show clearly maximum and minimum  possible   $M$ and $V_{\rm c}$
%at given radii.
Later, we investigate the models with different {\it mean} $i$
to infer the minimum and maximum possible masses of the dark matter halo of 
the SMC within the central 3 kpc. 

It is clear that the observed profiles are quite similar to
the profiles for the total component 
(i.e., the combination of stars and gas), though
$M$ and $V_{\rm c}$ around $R=1-2$ kpc for the total component are slightly
smaller than the observed ones.
Clearly, the observations can be well reproduced
without invoking the dark matter halo for the optical radius
of the SMC.
There is only a slight difference between the total mass ($M_{\rm t}$)
and the dynamical mass estimated from $V_{\rm c}$ within the central
3 kpc. If this difference is due to the presence of the dark matter halo,
the total mass ($M_{\rm dm}$) and the mean density (${\rho}_{\rm dm}$)
within the central 3 kpc are
$7.5 \times 10^7 {\rm M}_{\odot}$
and $6.7 \times 10^{-4} {\rm M}_{\odot}$ pc$^{-3}$,
respectively.

What would be a reasonable value for $M_{\rm s}/L_{\rm V}$ based on
integrated optical colors?
We use the formular described by Bell \& de Jong (2001, BD01)  
and the observed colors of the SMC: $B-V$ (=0.41) and $B-R$
(=0.70). 
If we use the low metallicity models with $Z=0.008$ in BD01,
the most plausible $M_{\rm s}/L_{\rm V}$ is about 1.
However, $M_{\rm s}/L_{\rm V}$ can range from 0.6 to 1.0 for different stellar population
models and star formation histories (BD01, S04). 
The stellar population synthesis models for ${\rm [Fe/H]} = -0.68$
and ages of 2, 10, and 13 Gyr by Vazdekis et al. (1996) 
show $M_{\rm s}/L_{\rm V} =1.1$, 3.8, and 4.4 respectively.
This means that $M_{\rm s}/L_{\rm V}$ strongly depends on  
the luminosity-weighted age of the stellar populations.
As it is observationally not clear what are the dominant ages of stellar populations,
we assume that $M_{\rm s}/L_{\rm V}$ is a free parameter
in the present study.
Given that the SMC contain a significant fraction of older stellar
populations (Harris \& Zaritsky 2004),
we consider that the reasonable $M_{\rm s}/L_{\rm V}$ would be
$2-4$ for  the observed large fraction of old stars
in the SMC (e.g., Harris \& Zaritsky 2004).

\begin{figure}
\psfig{file=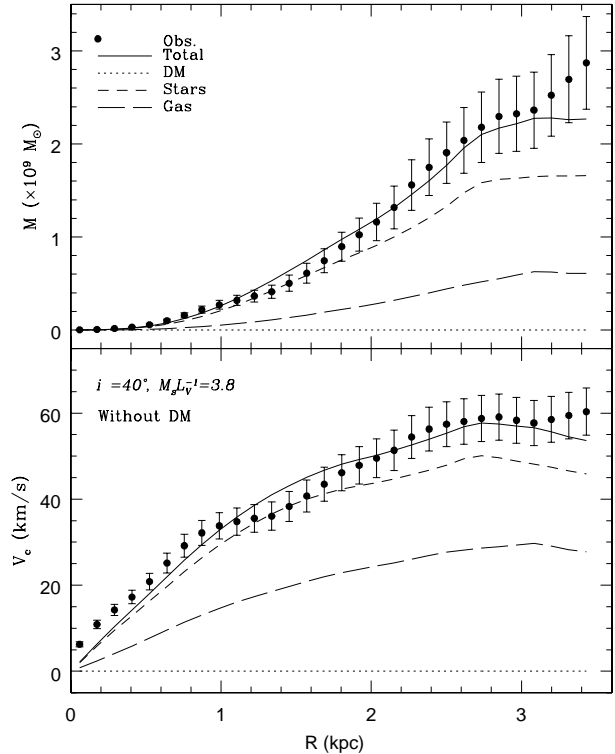,width=8.0cm}
\caption{The total masses ($M$, upper) within $R$
(where $R$ is the distance from the center of the SMC)
and circular velocities ($V_{\rm c}$, lower) at $R$
for all components (solid), dark matter (DM) one  (dotted),
stellar one (short-dashed), and gaseous one (long-dashed)
in the model NDM1 without dark matter.
Here the inclination angle $i$ and the $V$-band
stellar-to-mass-to-light-ratio ($M_{\rm s}/L_{\rm V}$)
are $40^{\circ}$ and 3.8, respectively. 
For comparison, the observed values are plotted by filled circles.}
\label{Figure. 1}
\end{figure}

\begin{figure}
\psfig{file=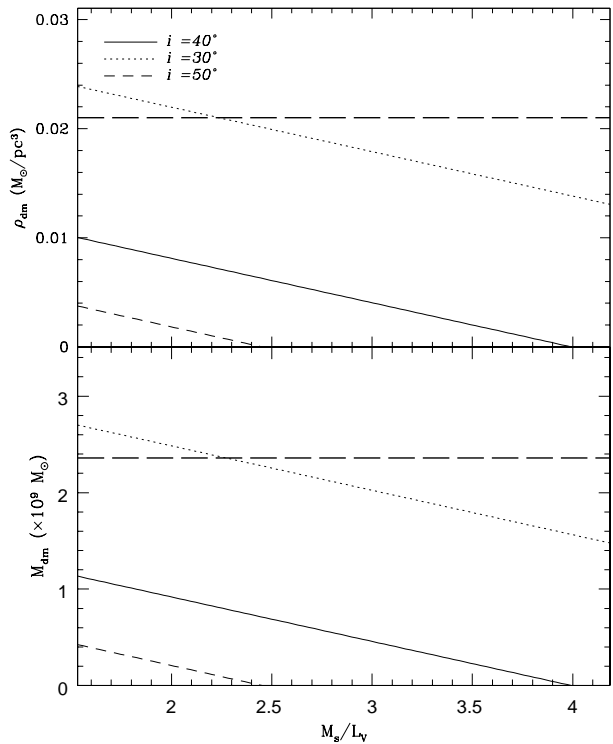,width=8.0cm}
\caption{The mean density  (${\rho}_{\rm dm}$, upper) 
and the total mass  ($M_{\rm dm}$, lower)
of the dark matter halos  within the central
3 kpc of the SMC as a function of $M_{\rm s}/L_{\rm V}$,
for $i=40^{\circ}$ (solid),
$i=30^{\circ}$ (dotted),
and $i=50^{\circ}$ (dashed). For comparison, 
the mean density (${\rho}_{\rm dm}$)
and the total mass ($M_{\rm b}$) for the baryonic
component of the SMC 
for $M_{\rm s}/L_{\rm V}=3.8$
are shown by dashed lines.
This figure indicates the possible upper and lower masses of the SMC
dark matter halo.}
\label{Figure. 2}
\end{figure}

\subsection{Uncertainties in $M_{\rm s}/L_{\rm V}$ and $i$}

We now explore how $M_{\rm s}/L_{\rm V}$ and $i$ affect 
$M_{\rm dm}$ and ${\rho}_{\rm dm}$.
Generally, for a given  $M_{\rm s}/L_{\rm V}$, $i<40$ degrees 
results in significantly larger $M_{\rm dm}$ and ${\rho}_{\rm dm}$.
We first calculate $M_{\rm dm}$ from  the total mass
(derived from the observed $V_{\rm c}$ at $R=3$ kpc
for a given $i$) and
and the observed gaseous  ($M_{\rm g}$)
and stellar masses ($M_{\rm s}$) for a given $M_{\rm s}/L_{\rm V}$:
$M_{\rm dm}=M-M_{\rm g}-M_{\rm s}$ so that no cold and warm gas 
possibly in the SMC's halo can be  considered.
We then estimate ${\rho}_{\rm dm}$ from the derived  $M_{\rm dm}$
within the central 3 kpc of the SMC.
Fig. 2 shows that 
$M_{\rm dm}$ and ${\rho}_{\rm dm}$ are significantly smaller 
than $M_{\rm b}$ (i.e., the total mass of the baryonic components
including both gas and stars)
and ${\rho}_{\rm b}$ (the mean density of the baryonic components),
respectively, for a reasonable $i$ ($=40^{\circ}$)
and $1.5 \le M_{\rm s}/L_{\rm V} \le 3.8$.
Only if $i$ is largely underestimated by $\sim 10^{\circ}$ 
in the observation by S04,
$M_{\rm dm}$ and ${\rho}_{\rm dm}$ can be 
significantly large: the model NDM2 with $i=30^{\circ}$
shows $M_{\rm dm}=1.6 \times 10^9 {\rm M}_{\odot}$
and ${\rho}_{\rm dm}=0.015 {\rm M}_{\odot}$ pc$^{-3}$.
Since $i$ is determined by
the tilted ring algorithm  at each $R$ in S04,
it is  unlikely that $i$ is underestimated 
by  $\sim 10^{\circ}$ {\it for the entire region}  of the SMC.
We thus conclude  that the larger $M_{\rm dm}$
due to the smaller $i$ is highly unlikely.

It should be stressed here that (i) the derived total mass
of $2.4 \times 10^9 {\rm M}_{\odot}$ for the best $i$ ($=40^{\circ}$) 
in S04 is consistent with that derived from stellar kinematics of the SMC
(e.g., Harris \& Zaritsky 2006) and (ii) the best $i$ is also
consistent with that for the best theoretical  model that can explain
physical properties of the Magellanic Stream and Bridge 
(e.g., Yoshizawa \& Noguchi 2003).  These strongly suggest that
although $i$ and position angles
change with $R$ in S04,  the best $i$ and position angle in S04
can be very close to the true ones.
This implies that the adopted tilted-ring algorithm in S04
is still useful in accurately estimating the rotation curve
for the inner part of the SMC ($R<3$  kpc), where the influences
of the  LMC-SMC-Galaxy interaction can be relatively weak.

The model NDM3 with  $M_{\rm s}/L_{\rm V} =2.3$
shows 
$M_{\rm dm} =7.8 \times  10^8 {\rm M}_{\odot}$
and ${\rho}_{\rm dm} =6.9 \times 10^{-3} {\rm M}_{\odot}$ pc$^{-3}$.
$M_{\rm dm}$ and ${\rho}_{\rm dm}$ are smaller for a larger  $M_{\rm s}/L_{\rm V} $:
$M_{\rm dm}$ and ${\rho}_{\rm dm}$ are $ 4.3 \times  10^8 {\rm M}_{\odot}$
and $ 3.8 \times 10^{-3} {\rm M}_{\odot}$ pc$^{-3}$,
respectively, for $M_{\rm s}/L_{\rm V} =3.1$.
The total mass of the dark matter halo within the optical radius of the SMC in the model NDM3 
 %$M_{\rm s}/L_{\rm V}$
is, however,  still low, given that the total baryonic
mass of the SMC is $1.6 \times 10^9 {\rm M}_{\odot}$.
The SMC may well be a baryon-dominated dwarf galaxy.
These results imply that only if the SMC 
%has a young age 
is dominated by a very young stellar population, 
and thus has a small $M_{\rm s}/L_{\rm V}$, the SMC can have a certain 
amount ($\sim 10^9 {\rm M}_{\odot}$) of dark matter  within its optical radius.

As a conclusion, the SMC can contain a certain amount of dark matter within its optical radius, 
if it has a significant fraction of young stellar populations and thus 
a small $M_{\rm s}/L_{\rm V}$ ($\sim 0.6$): 
$M_{\rm dm}$ and ${\rho}_{\rm dm}$ depend strongly on
$M_{\rm s}/L_{\rm V}$ such that they are both
larger for smaller $M_{\rm s}/L_{\rm V}$ for a given $i$.
The SMC can hardly contain dark matter, if it is dominated
by old stellar populations with
ages larger than 10 Gyr and thus has $M_{\rm s}/L_{\rm V} \ge 4$.
In the next section, we discuss possible mass distributions for the SMC's dark matter halo
that can be fitted to the observed 
$V_{\rm c}$ profile by assuming  $M_{\rm s}/L_{\rm V} =2.3$
(corresponding to ages of stellar populations similar to 5 Gyr)
that would be reasonable for a galaxy having a large fraction of  older stellar
populations like the SMC.
 
\subsection{Limitations of the mass estimation}

We have so far considered that the observed rotation curve
is very close to the true one with some possible
uncertainty of the disk inclination ($i$). 
It should be here stressed  that the observed rotation
curve can be slightly different from the true one
owing to the possible non-circular flow of gas caused
by the presence of a stellar bar in the SMC.
The latest survey of 2046 red giant stars has suggested
that the older stellar components
of the SMC have a velocity dispersion ($\sigma$)
of $\sim 27.5$ km s$^{-1}$
and a  maximum possible rotation 
of $\sim 17$ km s$^{-1}$
(Harris \& Zaritsky 2006).
The older stars are observed to have a more  regular distribution 
and appear  to be a slightly flattened ellipsoid
rather than a disk
(e.g., Cioni et al. 2000).
These  implies that the SMC has a stellar spheroid
rather than a rotating disk: the SMC is unlikely to have
a  bar composed of older stars.
The  lack of a stellar bar would not cause the non-circular
motion of gas in the SMC and thus suggests  that the observed
rotation curve by S04 is highly unlikely to be influenced
by the non-circular motion of gas due to the bar.

As discussed by S04,
numerous expanding shells in the SMC can  provide some non-circular motions 
and thus be  the main reason for the possible non-perfect rotation 
curve of the SMC. S04 however considered
this effect (i.e., introduction of velocity  dispersion) by 
calculating the asymmetric drift correction.
The last LMC-SMC tidal interaction about 0.2 Gyr 
can give tidal perturbation
to the SMC and thus influence gas dynamics of the SMC in its outer part
(Bekki \& Chiba 2007). Although the  outer part of gas ($3<R<5$ kpc) 
can be stripped to form the Magellanic Bridge during the interaction
(e.g., Yoshizawa \&
Noguchi 2003; Muller \& Bekki 2007),
the inner part of the present SMC 
(i.e., 0.2 Gyr after the last interaction)  can still show clear rotation.
Therefore, if the disordered gas motion
due to the last LMC-SMC interaction
can influence the present mass estimation of the SMC  from the observed
rotation curve,
the influence would be significantly small: a possible  influence
would be slight underestimation of the mass due to $V_{\rm c}$ smaller
than the true value resulted from a higher velocity dispersion
(caused by the last interaction)
in the outskirt of the SMC.

The 3D structure (e.g., thickness and shape) of 
the HI gas in the SMC remains observationally unclear. 
We thus can not make robust conclusions on how the thickness and shape
of the HI can influence the mass estimation of the SMC.
S04 adopted a reasonable assumption on a constant vertical scale-height (1kpc)
and considered  asymmetric drift corrections  in deriving the rotation
curve of the SMC. We here suggest that  a possible 
large intrinsic ellipticity (e.g., $\epsilon \sim 0.3$)
of the HI distribution can introduce
an observational error in the estimation of the rotation curve in the SMC.
Since the possible  range of $\epsilon$ for the SMC is observationally
unclear,  it is hard for the present study to make a quantitative
estimation for the possible observational error.
Keeping these limitations in mind, we discuss the models for the dark matter 
halo of the SMC in the following sections.

\begin{figure}
\psfig{file=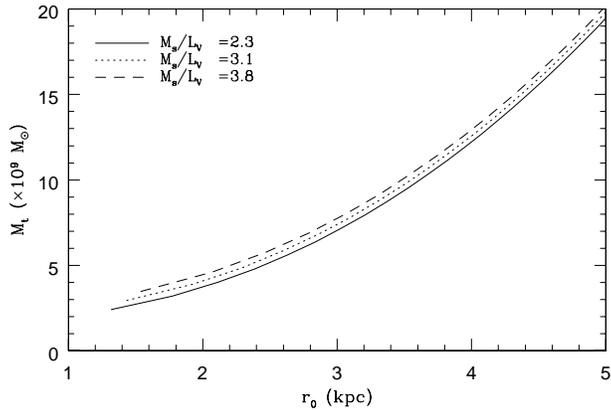,width=8.0cm}
\caption{
The relation between the core radius ($r_{0}$) of the 
SMC's dark matter halo
described as the Burkert profile (B95) and 
the total mass of the SMC ($M_{\rm t}$)
for $M_{\rm s}/L_{\rm V}=2.3$ (solid),
$M_{\rm s}/L_{\rm V}=3.1$ (dotted),
and  $M_{\rm s}/L_{\rm V}=3.8$ (dashed).
}
\label{Figure. 3}
\end{figure}

\begin{figure}
\psfig{file=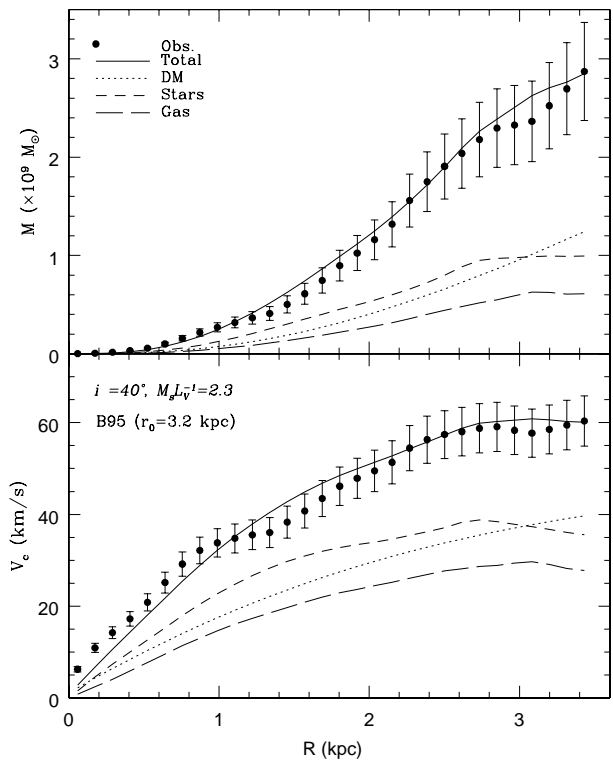,width=8.0cm}
\caption{
The same as Fig. 1 but for the model B1 in which the Burkert
profile with $r_0 = 3.2$ kpc, $i=40^{\circ}$, 
and $M_{\rm s}/L_{\rm V}=2.3$  are adopted.
$M_{\rm dm}$ and $M_{\rm t}$ are 
$ 6.5 \times 10^9 {\rm M}_{\odot}$ and
$ 8.1 \times 10^9 {\rm M}_{\odot}$, respectively, in this model.
Note that the Burkert profile with a large core radius can well reproduce
the slowly rising $V_{\rm c}$ profile.
}
\label{Figure. 4}
\end{figure}

\begin{figure}
\psfig{file=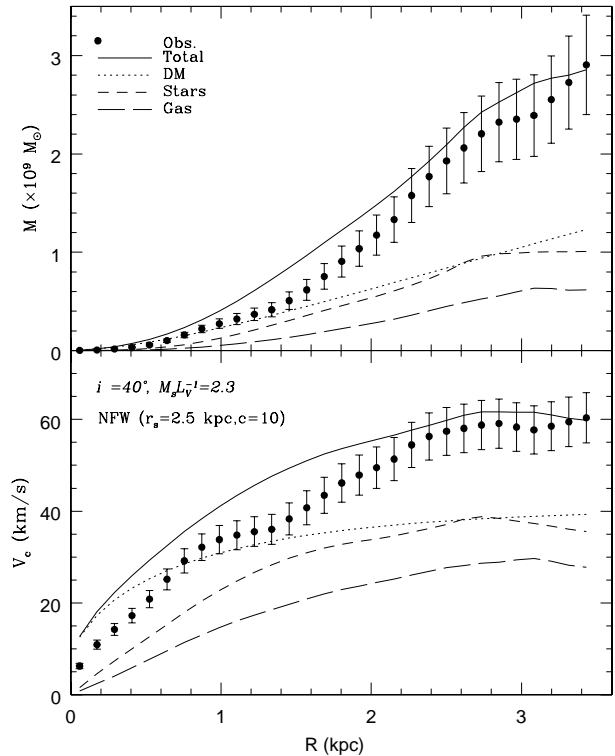,width=8.0cm}
\caption{
The same as Fig. 1 but for the model NFW1 in which the NFW 
profile with $r_{\rm s} = 2.5$ kpc and $c=10$,  $i=40^{\circ}$, 
and $M_{\rm s}/L_{\rm V}=2.3$  are adopted.
$M_{\rm dm}$ and $M_{\rm t}$ are 
exactly the same as those used in the Burkert one
shown in Fig. 4.
Note that the fit of the NFW profile 
to the observation is worse 
than the Burkert profile in Fig. 4,
though the NFW profile can better reproduce the  
$V_{\rm c}$ profile in the outer part ($R \sim 3$ kpc).
}
\label{Figure. 5}
\end{figure}

\begin{figure}
\psfig{file=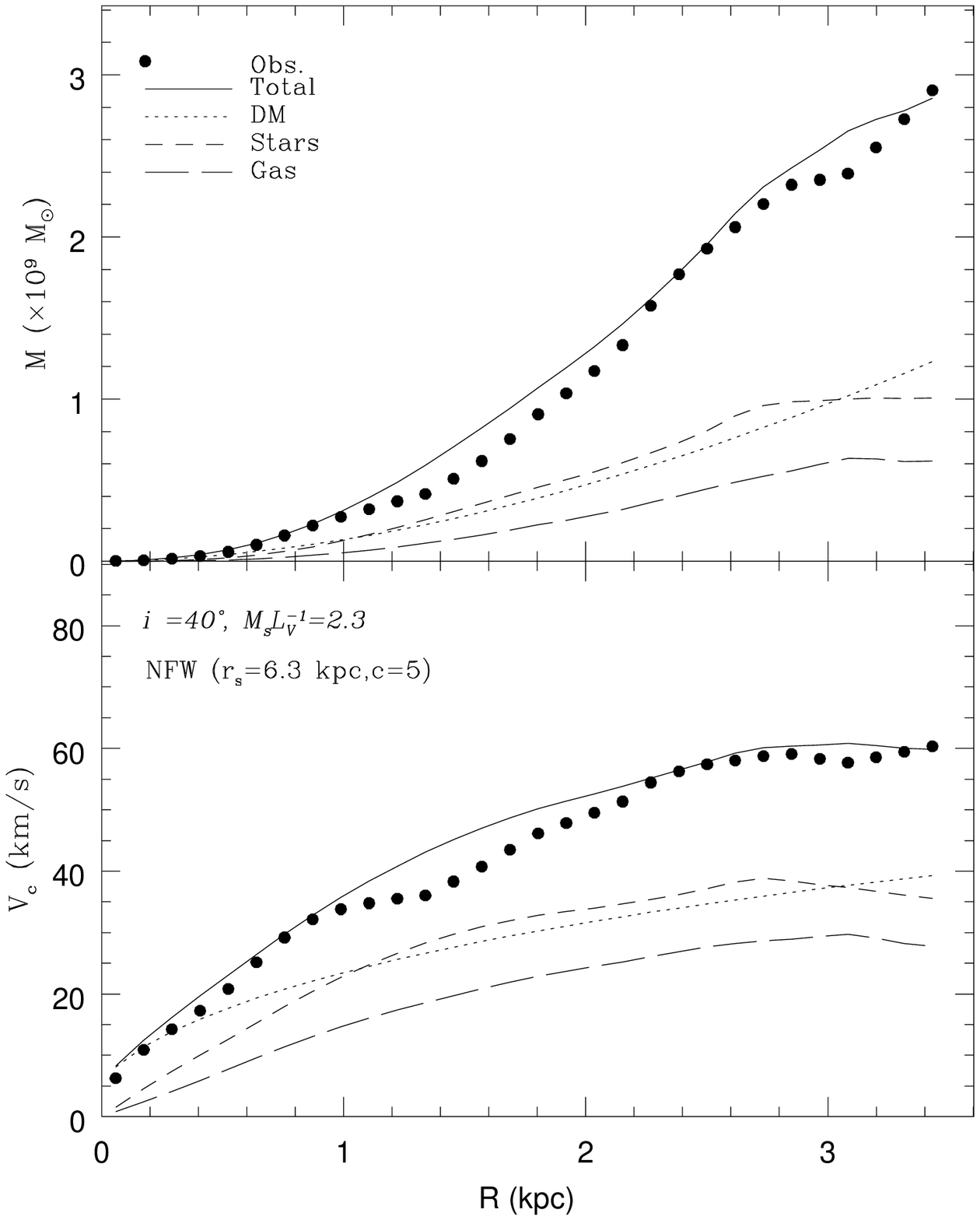,width=8.0cm}
\caption{
The same as Fig. 1 but for the model NFW3 in which the NFW 
profile with $r_{\rm s} = 17.7$ kpc and  $c=5$,  $i=40^{\circ}$, 
and $M_{\rm s}/L_{\rm V}=2.3$  are adopted.
In this model,  unusually  large mass and low density
of the dark matter halo for the SMC are assumed
just for comparison with the results shown in Fig. 4 and 5.
}
\label{Figure. 6}
\end{figure}

\section{The models for the dark matter halo}

\subsection{The NFW profile}

We now investigate the $M$- and $V_{\rm c}$-profiles
in the models with dark matter halos and $M_{\rm s}/L_{\rm V}=2.3$
reasonable for the SMC having a large fraction of intermediate-age
and old stellar populations. 
We adopt the same radial density profiles (described later)
of dark matter halos as widely adopted in other studies
(on isolated systems with no apparent  galaxy interaction),
though the SMC appears to be interacting with the LMC and the Galaxy.
We think that this adoption can be reasonable, because we discuss
the density profile in the inner part of the dark halo, where
the influence of the Galaxy-LMC-SMC interaction is relatively small.

We adopt the following two profiles for the 
radial density distribution of the dark matter
halo of the SMC: the ``NFW'' profile 
with a  central cusp predicted
from hierarchical clustering scenarios based on
the cold dark matter models (Navarro et al. 1995)
and the ``Burkert'' profile with a flat ``core'' (Burkert 1995; B95).

The NFW profile is described as:
\begin{equation}
{\rho}_{\rm dm}(r)=\frac{\rho_{0}}{(r/r_{\rm s})(1+r/r_{\rm s})^2},
\end{equation}
where $r$,  $\rho_{0}$,  and $r_{\rm s}$ are the distance from the
center
of the cluster, the scale density, and the scale-length of the dark
halo,
respectively. The ratio of the virial radius ($r_{\rm vir}$)
and $r_{\rm s}$ is a parameter in the NFW profile
and described as $c$.
A dark matter halo with the total mass of
$6.5 \times 10^9 {\rm M}_{\odot}$ and $c=10$
has  $r_{\rm s}=2.5$ kpc and the virial velocity
of $33$ km s$^{-1}$ in the NFW profile.
The total mass in a NFW  model is defined as the mass within
$r_{\rm vir}$ in the  present study.

\subsection{The Burkert profile}

The Burkert profile is described as:
\begin{equation}
{\rho}_{\rm dm}(r)=\frac{\rho_{dm,0}}{(r+r_{0})(r^2+{r_{0}
}^2)},
\end{equation}
where $\rho_{dm,0}$ and $r_{0}$ are the central dark matter
density and the core (scale) radius, respectively.
In the Burkert profile (B95), 
a relation between
the total mass within the core ($M_{0}$) and $r_{0}$
is  described as:
\begin{equation}
M_{0}=7.2 \times 10^7 {(\frac{ r_{0} }{ 1 kpc })}^{-7/3}  {\rm M}_{\odot}.
\end{equation}
The virial radius ($R_{\rm h}$)
and the total mass of the dark matter halo in this profile
are  equal to $3.4 \times r_0$ and $5.8 \times M_{0}$, respectively.
Therefore the total mass of the dark matter halo 
within $R_{\rm h}$ in the SMC
can determine the core radius ($r_0$).
Fig. 3 shows a relation between $r_0$ and $M_{\rm t}$
($=M_{\rm dm}+M_{\rm b}$) derived from the equation (3)
for three different $M_{\rm s}/L_{\rm V}$.
The  parameter values (e.g., $r_{\rm s}$ and $r_0$)
of models that are investigated in detail
are summarized in the Table 1.

The tidal radius of the SMC is about 5 kpc (Gardiner \& Noguchi 1996)
so that the shape of the dark matter halo can keep its (assumed)
spherical shape within  the central 3 kpc after the LMC-SMC
interaction.  Therefore, the assumed spherically symmetric 
distribution of the dark matter halo in both the NFW and Burkert
profiles can be regarded as reasonable even for the SMC 
about 0.2 Gyr after the last strong interaction with the LMC.

\subsection{Comparison with observations}

Fig. 4 shows  that the model B1 having the Burkert profile with
$M_{\rm dm}=6.5 \times 10^9 {\rm M}_{\odot}$ and
$r_0=3.2$ kpc (i.e., $M_{\rm t}=8.1 \times 10^9 {\rm M}_{\odot}$
and $R_{\rm h}=11.0$ kpc) can well reproduce
observations in terms of
both (i) the slowly  rising $V_{\rm c}$ profile
in the inner part ($R<2$ kpc) and (ii) the maximum  $V_{\rm c}$
($\sim 60$ km s$^{-1}$).
The total mass of the dark matter halo within 3 kpc is about
$10^9 {\rm M}_{\odot}$, which is still smaller than
the total baryonic mass.
These results imply that the SMC can have a large core in
the mass distribution of the dark matter halo.
We confirm that more massive models (e.g., B2) 
with larger $r_0$ can also reproduce well 
the observed $V_{\rm c}$ profile.
The results of the models with lower $M_{\rm s}/L_{\rm V}$ (=0.6 and 1)
are shown in the Appendix A for comparison.

Fig. 5 shows the simulated  $V_{\rm c}$ as a function of $R$ 
for the model NFW1 having the NFW profile with  $c=10$, $r_{\rm s}=2.5$,
and $M_{\rm dm}$ being exactly the same
as that used in the Burker profile shown in Fig. 4.
It is clear that although the NFW profile can reproduce the 
maximum  $V_{\rm c}$ around $R\sim3$ kpc, it can not fit
to the observed $V_{\rm c}$ profile as well as the Burkert one.
In particular, in the central region ($R<2$ kpc) the simulated profile
is systematically  well above the observed data points. 
The higher mass-density in the inner region of the dark matter 
halo described in the NFW profile is responsible for
this discrepancy.   
These results imply that the NFW profile 
with reasonable model parameters are highly
unlikely to reproduce the observations.

It should be however stressed that au unusually  low-density
(i.e., smaller $c$), high-mass dark matter models with the NFW profile
can reproduce the observations as well as the Burkert profiles.
For example, the simulated $V_{\rm c}$  profile in the model NFW2 with  
$M_{\rm dm}=6.5 \times 10^9 {\rm M}_{\odot}$,
$c=5$, and $r_{\rm s}=5.1$ kpc 
can be well consistent with the observations.
Fig. 6 shows the simulated $V_{\rm c}$  profiles
for the model NFW3 with
$M_{\rm dm}=7.9 \times 10^{10} {\rm M}_{\odot}$,
$c=5$, and $r_{\rm s}=17.7$ kpc.
In this extreme example with a low-density dark matter halo,
the observations can be well reproduced,
although it is unlikely that the mass of the SMC is
much larger than that of the LMC ($\sim 2 \times 10^{10} {\rm
M}_{\odot}$) that was previously  considered to form a binary
with the SMC (e.g., Murai \& Fujimoto 1980).

These results imply that the Burkert profile
is better than the NFW one in terms of reproducing
HI rotation curve in the SMC.
Although these results do not rule out the NFW profile
for the distribution of the dark matter halo of the SMC,
the observed low-density dark matter halo
(or large central concentration of the baryonic components)
needs to be explained in the context
of hierarchical clustering scenarios.
If, however, the SMC has  such a large initial mass of 
$M_{\rm dm}=7.9 \times 10^{10} {\rm M}_{\odot}$,
as shown in Fig. 6, then our views of the formation and evolution processes in 
the Magellanic system can be dramatically changed.

\begin{figure}
\psfig{file=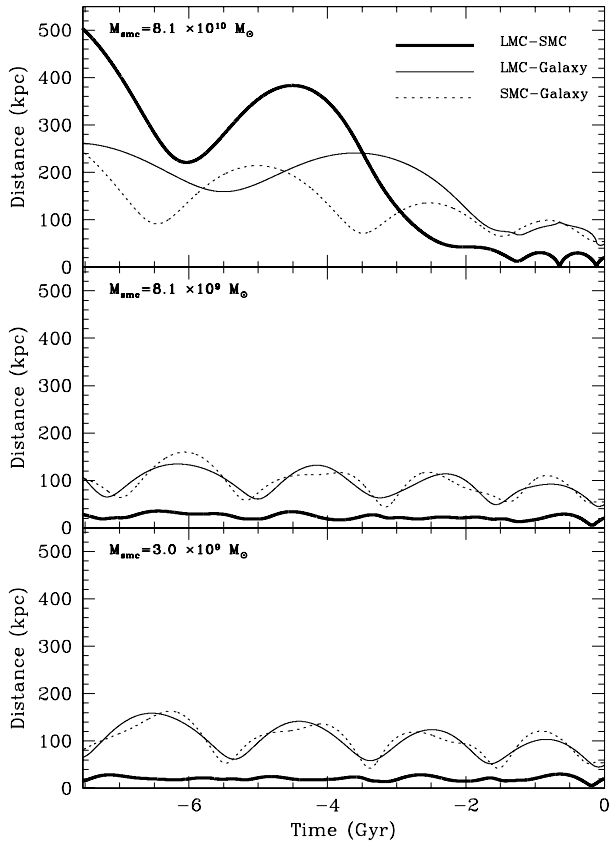,width=8.0cm}
\caption{
Orbital evolution of the MCs for
$M_{\rm smc}=8.1\times 10^{10} {\rm M}_{\odot}$ (top),
$M_{\rm smc}=8.1 \times 10^9 {\rm M}_{\odot}$ (middle), and
$M_{\rm smc}=3.0 \times 10^9 {\rm M}_{\odot}$ (bottom).
The time evolution of distances between MCs
(thick solid), the LMC and the Galaxy (thin solid),
and the SMC and the Galaxy (dotted) are separately shown.
Here the total mass of the LMC 
is assumed to be $2.0 \times 10^{10} {\rm M}_{\odot}$.
}
\label{Figure. 7}
\end{figure}

\begin{figure}
\psfig{file=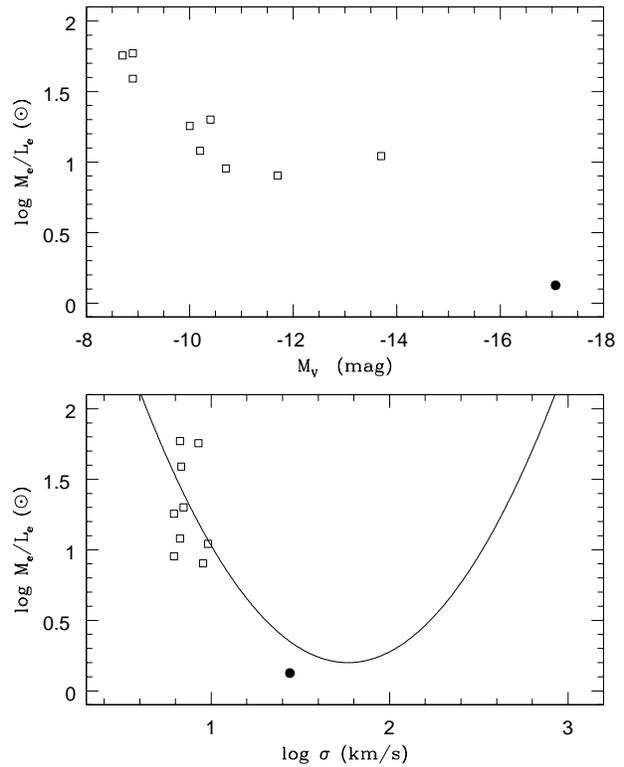,width=8.0cm}
\caption{
The dependences of $M_{\rm e}/L_{\rm V}$ on $M_{\rm V}$
(upper) and on $\sigma$ (lower) within effective radii
for the dwarfs in the Local group (open squares) and
the SMC (filled circles). 
Here $M_{\rm e}$ and  $\sigma$ represent the total masses
(including dark matter halos)
within effective radii and velocity dispersions for
dwarfs, respectively. 
Here the observational
results by  Mateo (1997) are used and the solid line
in the lower panel is from Zaritsky  et al. (2006).
}
\label{Figure. 8}
\end{figure}

\section{Discussion}

\subsection{A  massive past of the SMC} 

If the SMC has low-mass and low-density of the dark matter,
as shown in our models, 
then the present results have the following four implications.
Firstly, the SMC could have lost most of its original mass through tidal interactions
with the Galaxy and the LMC owing to the extended initial
distribution of the dark matter halo.
Essentially, the present-day SMC is  a stripped
core of an initially much larger galaxy.
The SMC's tidal radius ($r_{\rm t}$)
against the Galaxy can range from 5 kpc to 7.5 kpc
for the total mass ($M_{\rm t}$)
of $3-10 \times 10^9 {\rm M}_{\odot}$ (see the formula
for the mass estimate in Gardiner \& Noguchi 1995).
Therefore  $R_{\rm h}$ (11 kpc) for
the SMC with $M_{\rm dm}=6.5 \times  10^9 {\rm M}_{\odot}$
is significantly  larger than $r_{\rm t}$,
which means that the SMC could have lost all components outside
$r_{\rm t}$ owing to tidal stripping.
The stripped matters may well be now observed as the MS, the MB,
and the stream of dark matter in the outer halo of the Galaxy.

Secondly, the SMC could have much more strongly influenced
the LMC owing to its larger mass, in particular, 
for the last  3 Gyrs.  The previous models 
assumed that $M_{\rm t} \sim 3 \times 10^9 {\rm M}_{\odot}$
(e.g., Gardiner \& Noguchi 1995; Yoshizawa \& Noguchi 2003;
Bekki \& Chiba 2005, BC05),
which thus would have underestimated gravitational influences of the SMC
on the LMC. Thirdly, the SMC may well
initially  have a larger amount of HI gas  
owing to its large total mass.  The total gas mass of the SMC
in previous MS models is limited to less than $10^9 {\rm M}_{\odot}$
owing to the adopted models with smaller $M_{\rm t} $ (
e.g., Yoshizawa \& Noguchi 2003).
This limitation was responsible for the less consistent model
for the total mass of the MS and the MB (Yoshizawa \& Noguchi 2003).
Thus if future theoretical works  adopt  a more massive SMC model, 
they will much more consistently explain the observed total
mass ($>10^9 {\rm M}_{\odot}$) of the MS and the MB.

Fourthly,  future orbital evolution models for the MCs
would need to be revised significantly  by considering
the possible large total mass of the SMC. Since the orbital evolution of the MCs,
which strongly depends on the mass of the SMC,
is the most important ingredient for the formation of the MS and the MB
(e.g., Yoshizawa \& Noguchi 2003; Muller \& Bekki 2007),
the models with larger $M_{\rm t} $ can describe the formation processes
in a quite different manner: future theoretical works  on the MS and MB
formation need to investigate models with a larger initial
mass of the SMC.

In order to illustrate how the orbital evolution of the MCs
can be changed owing to a larger mass of the SMC,
we investigate the evolution based on the same orbital evolution model
used in previous works (e.g., BC05): the adopted initial
parameters on  initial positions
and velocities of the MCs and the gravitational potential of 
the Galaxy are exactly the same as those in BC05.
Fig. 7 shows the orbital evolution of the MCs
for the models in which (i) the methods to model
orbital evolution of the MCs are the same as  those used in 
BC05 as  ``no dynamical friction'',  
and (ii) $M_{\rm t}$ ($=M_{\rm smc}$)
is equal to or much larger than that 
used in BC05 ($3 \times 10^9 {\rm M}_{\odot}$). 
It is clear from this figure that even if dynamical friction between the MCs
is not included, the MCs can not keep their binary status
for more than a few Gyr in the model with large $M_{\rm t}$. 
We confirm that if the SMC has  $M_{\rm t} > 1.5 \times 10^{10} {\rm M}_{\odot}$,
the MC can not keep its binary status for more than 4 Gyrs.

If dynamical friction between the MCs can be effective,
the MCs can not keep their binary status for more than 4 Gyrs
even for $M_{\rm t}=3 \times 10^{10} {\rm M}_{\odot}$
(BC05): the more massive SMC suggested in the present study
is highly unlikely  to be coupled with the LMC for
more than 4 Gyrs. Furthermore,  the evolution processes (e.g., star formation
history) of the SMC can be significantly  different in the models with larger  $M_{\rm t} $
owing to the stronger self-gravity  of the SMC against
tidal perturbations from the Galaxy and the LMC. 
Thus HI observations on the mass distribution of the SMC
(e.g., S04) may well  change significantly views about
formation and evolution of the MCs.

Sofue (1998) analyzed the rotation curve profile of M82 and found
that the profile declines in a Keplerian fashion
outside 200 pc.
He thus suggested that (i) a significant amount of mass is
missing in the outer part of M82 and 
(ii) tidal stripping due to past tidal interaction
between M81 and M82 is responsible for the missing mass.
These results by Sofue (1998) and the present ones suggest
that the SMC would be  the second case of low-mass galaxies that
have lost a significant fraction of their original masses.
A significant  difference between these two galaxies
is that the rotation curve of the SMC is not observed to show 
a decline in a Keplerian fashion even for the outer part.

\subsection{The closest dIrr or dSph  with a flat DM core} 

The present study has first suggested that
the dark matter halo of the SMC  has a flat core 
well described by the Burkert profile rather than the NFW one:
the SMC can be  the nearest example of dIrr's (or dSph's) in which
the NFW profiles are hard to fit to HI rotation curves.
It should be however stressed that Valenzuela et al. (2007) have recently 
suggested a possibility of a large underestimate of 
the HI rotational velocity in the central 1 kpc region of two dwarf galaxies
due to noncircular motions.  Since Valenzuela et al. (2007) adopted
rotating disk models for their investigation,
their results can not be applied directly to the SMC which has no evidence of 
stellar rotation (Harris \& Zaritsky 2006).

However, if the last close LMC-SMC interaction (Bekki \& Chiba 2007)
could cause the radial flow of HI gas and thus the noncircular motions,
then the observed $V_{\rm c}$ in S04  could be underestimated.
If this is the case,  the NFW profile with a cuspy core
would be still possible. 
The present models are not based on numerical simulations
of HI gas in the SMC thus can not allow us to make it
clear whether 
the underestimation of circular velocities suggested
by Valenzuela et al. (2007) is equally possible in the gas disk
of the SMC interacting with the Galaxy and the LMC.
We plan to discuss this problem
in our future paper in which the HI rotation curve 
of the SMC will be constructed 
based on hydrodynamical simulations of the SMC evolution.

The derived small mass of the dark matter halo within the optical radius 
of the SMC suggests that the SMC is a baryon-dominated galaxy.
This strong central concentration of baryonic components in the SMC
also suggests that the formation history of the SMC can be significantly
different from that of other dIrr's dominated by dark matter halos.
Fig. 8 shows how unique the SMC is by comparing
the location of the SMC and dwarfs
in the Local group  on the $M_{\rm V}-M_{\rm e}/L_{\rm V}$ plane
and  $\sigma-M_{\rm e}/L_{\rm V}$ one,
where  $M_{\rm e}$ and  $\sigma$ represent the total masses
(including dark matter halos)
within effective radii and velocity dispersions, respectively.
Although the SMC has much lower $M_{\rm V}/M_{\rm e}$ 
in comparison with other dwarfs in the Local group,
it is possible that $M_{\rm V}/M_{\rm e}$ is not
so low for a luminous dwarf population
(i.e., more luminous dwarfs can have lower $M_{\rm V}/M_{\rm e}$;
see Zaritsky et al. 2006).

If the SMC really has a very low-density dark matter halo
and thus a strong concentration of its baryonic components
within the central 3kpc, then the origin of this unique nature of the SMC needs to be
understood. Efficient transfer of gas to the central regions
of galaxies and the resultant starbursts there in galaxy merging (e.g., Mihos  \& Hernquist 1995)
may significantly  increase the baryonic fractions of merger remnants.
Recently, Bekki \& Chiba (2008) have suggested that the SMC
was formed from merging of two gas-rich dIrr's with extended HI gas
very long time ago. Therefore, the observed large fraction of baryonic component
in the SMC can be  due mainly to the past major merger event
that formed the SMC. Thus the observed HI kinematics of the SMC
can provide unique insights into the formation history
of the SMC.

\subsection{Stronger constraints on the mass of the SMC from
future observations}

One of potential problems in estimating the mass of the SMC
based on the observed $V_{\rm c}$ profile is that we can underestimate significantly 
the SMC's inclination angle (e.g., by 20 degrees in $i$).
The HI distribution and kinematics are highly disturbed by numerous expanding shells (S04), 
making a more accurate estimate of the inclination angle very difficult.
A significant underestimate of the inclination angle, although highly unlikely,
would result in  a larger total mass of the SMC within the central 3 kpc  
(e.g., $5.3 \times 10^9 {\rm M}_{\odot}$  for the true $V_{\rm c}$ of 80  km s$^{-1}$).
%This possible underestimate of $i$, however, implies 
%that the SMC could be significantly  more massive than what was previously thought.
Therefore, future observations of the distribution and radial velocities of a large number 
of older stars (e.g., RGB and AGB stars)
%with the number of stars much larger than that used 
%in Harris \& Zaritsky (2006) 
are very important to derive the mass of the SMC independently from the HI observations (S04). 

Although the present study has shown that the SMC  can have a small amount
of dark matter within the central 3 kpc, the detailed profile of the dark matter distribution
in the SMC has not been determined.
The observed $V_{\rm c}$ profile within the central 3 kpc alone
does not allow us to distinguish between 
the Burkert model with $M_{\rm t} \sim 6.5 \times 10^9 {\rm M}_{\odot}$ 
and a very massive Burkert (or even NFW) profile with 
$M_{\rm t}>10^{10} {\rm M}_{\odot}$. 
The mass profile of the SMC at $R>3$ kpc,
where less massive models can show almost flat $V_{\rm c}$ profiles,
need to be observationally investigated to provide stronger  constraints on the distribution
of the dark matter halo in the SMC.

No\"el \& Gallart (2007) have recently found 
intermediate-age and old stars belonging to the SMC but being located
6.5 kpc from the SMC center.  This suggests that
the SMC stellar population may be more massive than previously thought.
%has the outer stellar halo component and is possibly more massive
%than other observations have ever suggested.
Future spectroscopic observations of radial velocities of the stars in the outer halo
will enable us to derive  the mass profile for the outer regions
of the SMC and thus to better constraint the radial mass profile
of the dark matter halo.
The ongoing VMC project (the VLT VISTA near-infrared
$YJK_{\rm s}$ survey for the MCs), combined with other large projects,
will soon provide structural and kinematical data sets for the outer
regions of the SMC (e.g., Cioni et al. 2007).
We plan to discuss the entire mass profile of the dark matter halo
of the SMC based both on these ongoing observations and previous
HI observations (S04) in our future papers.

\section{Conclusions}

We have investigated
the mass ($M_{\rm dm}$) and the mean density  (${\rho}_{\rm dm}$)
of the dark matter  halo within the optical radius ($\sim 3$ kpc) of
the SMC based on the high-resolution HI observations of the SMC. 
Although there are some limitations of the present models
in estimating the total mass of the SMC solely from
the observed HI rotation curve by S04,
we have tried to derive the  mass and best possible
dark matter models. 
The results and their implications are described as follows.

(1) The total mass, mean density, and  distribution of the dark matter halo
depend strongly on the adopted $M_{\rm s}/L_{\rm V}$.
For example,
the estimated $M_{\rm dm}$ and ${\rho}_{\rm dm}$ are
$7.5 \times 10^7 {\rm M}_{\odot}$
and $6.7 \times 10^{-4} {\rm M}_{\odot}$ pc$^{-3}$,
respectively, for 
$M_{\rm s}/L_{\rm V} =3.8$.
This implies that if the SMC is dominated by older stellar populations,
the SMC has a low-density dark matter halo described  by
B95 rather than by the NFW.

(2) The models with lower $M_{\rm s}/L_{\rm V}$ ($<1$) 
can show higher $M_{\rm dm}$ and ${\rho}_{\rm dm}$ 
yet have difficulties in reproducing the observed rotation curve,
in particular,  the inner part ($R<1$ kpc) of the curve,
both for the Burkert and the NFW profiles. 
This implies that the SMC is unlikely to be dominated by
younger stellar populations.

(3) The Burkert profile 
can fit better the observed HI rotation curve of the SMC 
than the NFW profile for a reasonable $M_{\rm dm}$ and $M_{\rm s}/L_{\rm V}$
for reasonable   $M_{\rm s}/L_{\rm V}$ ($\sim 2$).
However the NFW profiles with low-densities, small $c$ ($\sim 5$),
and large total masses ($M_{\rm dm} >   10^{10} {\rm M}_{\odot}$) 
can be also fit well the observations, though such low-density NFW profiles are highly unlikely
for low-mass dwarfs like the SMC. The SMC can be the closest example
among dIrr's in which rotation curves are  difficult to
fit to the mass profiles of dark matter halos predicted by
hierarchical clustering models.

(4) The dark matter halo of the SMC is likely to have the initial total mass 
of ${\rm M}_{\rm dm} \sim  6.5 \times 10^9 {\rm M}_{\odot}$
and the core radius of $r_{\rm 0} \sim 3.2$  kpc,
which means that the SMC could be much more massive than what was previously thought.

(5) Orbital evolution models of the SMC and the LMC,  on which formation models of the MS
and the MB  are based, would need to be  revised by considering the derived
much larger total mass of the SMC. The observed large concentration of the baryonic
components in the SMC is due to its unique formation history
and is very different from  those of other dIrr's.

\section*{Acknowledgments}
We are  grateful to the anonymous referee for valuable comments,
which contribute to improve the present paper.   
K.B. acknowledges the financial support of the Australian
Research.

\appendix

\section{The models with lower  mass-to-light-ratios}

The models with  lower $M_{\rm s}/L_{\rm V}$ (i.e.,
smaller stellar masses) have larger
masses of the dark matter halo and thus larger core (or scale) radii
for the SMC. It would be therefore interesting to discuss
whether such more massive models can explain
the {\it inner part} of the observed rotation curve.
Fig. A1 shows that for $M_{\rm s}/L_{\rm V}=0.6$,
even the very massive dark matter model (B3) with $r_0=9.5$ kpc
(corresponding to  $M_{\rm dm} =7.9 \times 10^{10} {\rm M}_{\odot}$)
with the Burkert profile can not explain the observed
profile of the rotation curve owing to the required large
mass in the inner region of the SMC.
Fig. A2 also shows that  for $M_{\rm s}/L_{\rm V}=1.0$,
the massive model (B4) can not reproduce the inner part
of the observed rotation curve owing to the low density of the halo.
It should be stressed here that the SMC is highly unlikely to have
a  mass much larger  than that of the LMC ($\sim 2 \times 10^{10} 
{\rm M}_{\odot}$.

These results imply that in order for the observed rotation curve
to be reproduced by the present model,
a higher $M_{\rm s}/L_{\rm V}$ needs to be adopted:
the SMC is highly unlikely to be dominated by younger stellar 
populations. This is consistent with the results by
Harris \& Zaritsky (2004) who showed that
about 50\% of the stellar populations in the SMC
were formed prior to 8.4 Gyr ago.
Given that the large fraction of stars in the SMC are suggested
to be formed between 2.5 and 3  Gyr ago in
starbursts (Harris \& Zaritsky 2004),
the SMC must be dominated by intermediate-age and old stellar
populations. Thus we can conclude that our
successful reproduction of observations in the models with
higher $M_{\rm s}/L_{\rm V}$ $\sim  2-4$ is quite reasonable.

\begin{figure}
\psfig{file=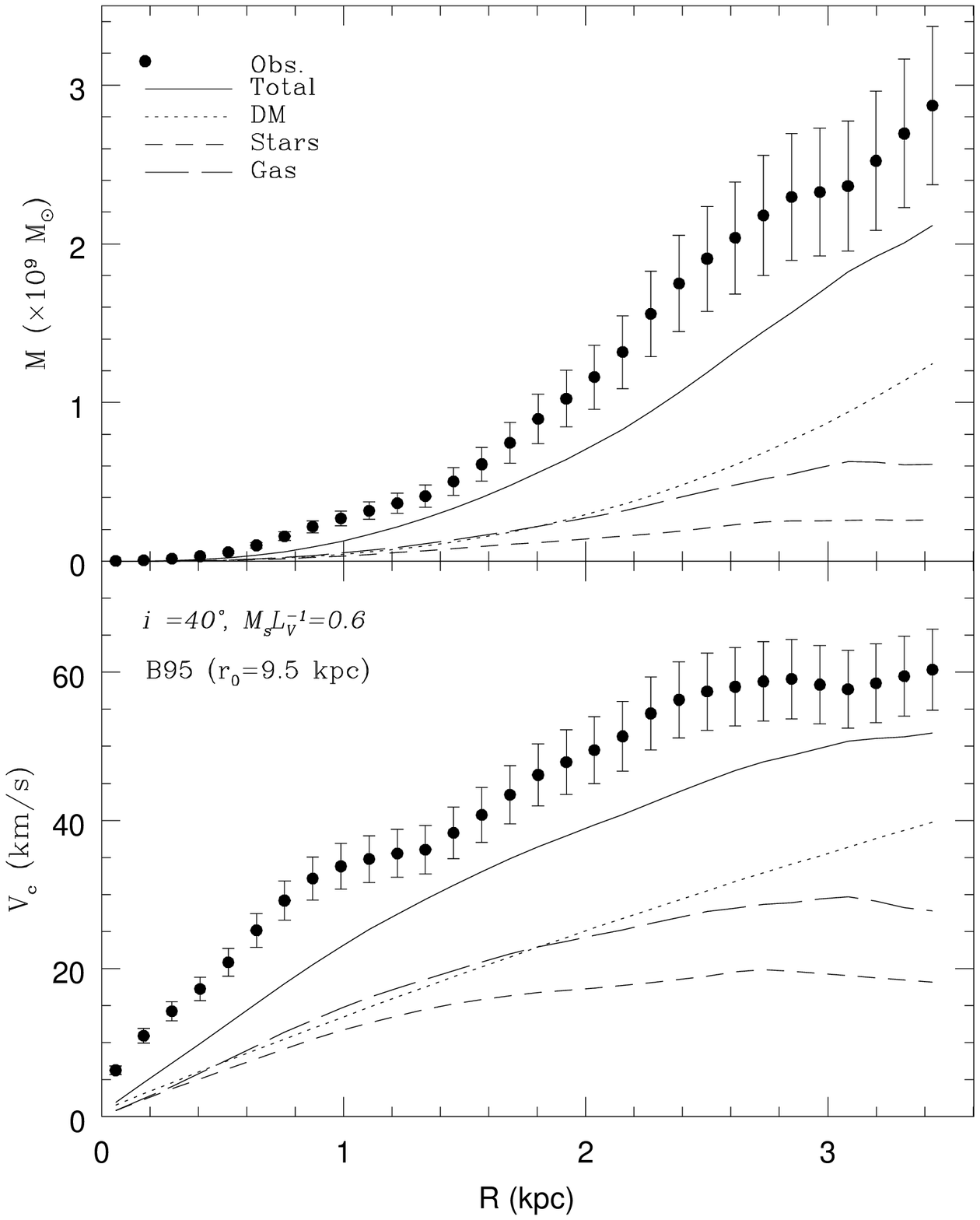,width=8.0cm}
\caption{
The same as Fig. 1 but for the model B3 in which the Burkert
profile with $r_0 = 9.5$ kpc, $i=40^{\circ}$, 
and $M_{\rm s}/L_{\rm V}=0.6$  are adopted
($M_{\rm dm} =7.9 \times 10^{10} {\rm M}_{\odot}$).
}
\label{Figure. 9}
\end{figure}

\begin{figure}
\psfig{file=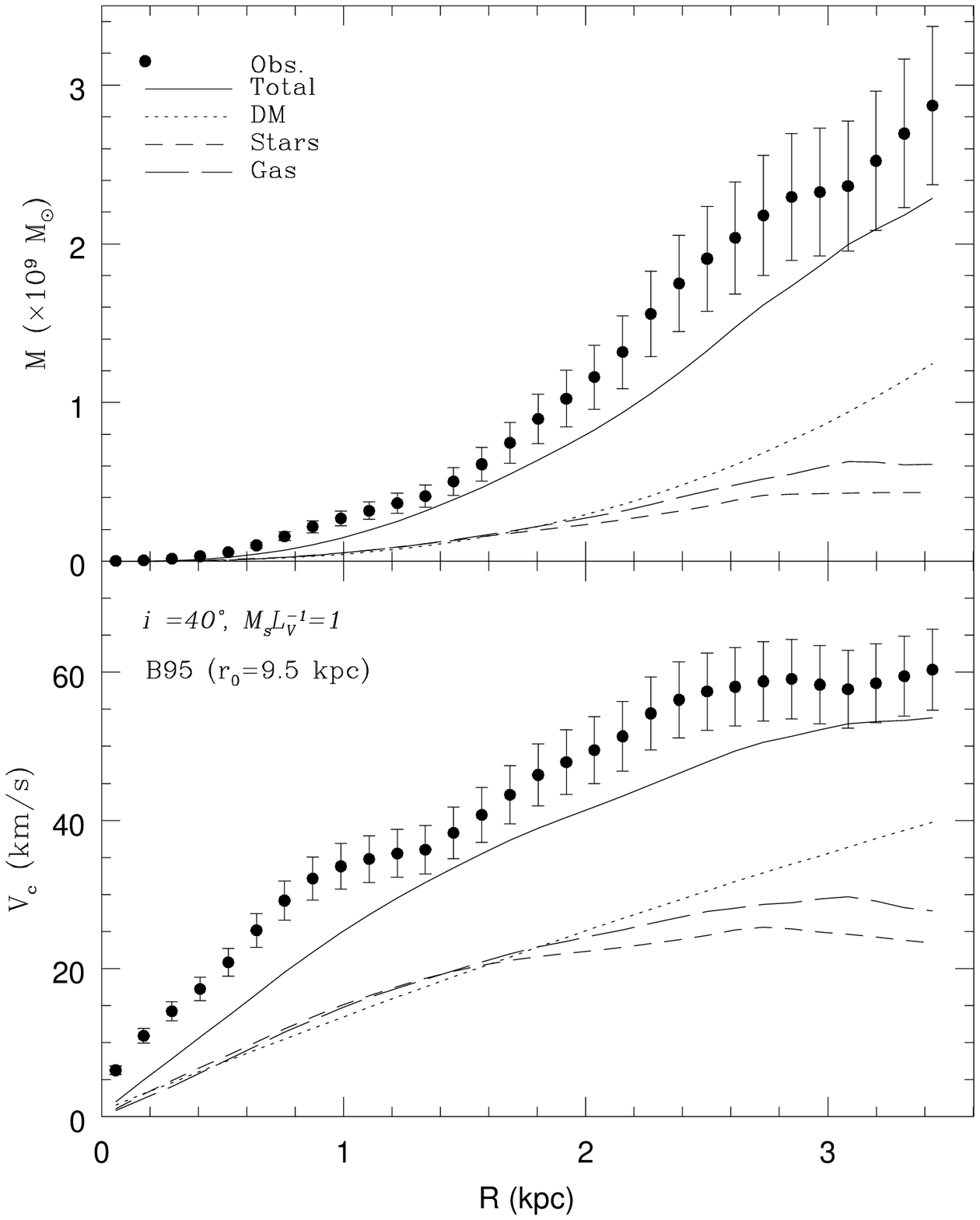,width=8.0cm}
\caption{
The same as Fig. 1 but for the model B4 in which the Burkert
profile with $r_0 = 9.5$ kpc, $i=40^{\circ}$, 
and $M_{\rm s}/L_{\rm V}=1$  are adopted
($M_{\rm dm} =7.9 \times 10^{10} {\rm M}_{\odot}$).
}
\label{Figure. 10}
\end{figure}
\end{document}